\documentclass[usenatbib]{mn2e}
\usepackage{graphicx}
\usepackage{picture}

\title[Stellar-remnant black holes in globular clusters]{Multiple stellar-mass black holes in globular clusters: theoretical confirmation}
\author[Anna C. Sippel, Jarrod R. Hurley]{Anna C. Sippel$^{1}$\thanks{E-mail: asippel@astro.swin.edu.au}, Jarrod R. Hurley$^{1}$\\
$^{1}$Centre for Astrophysics and Supercomputing, Swinburne University of Technology, PO Box 218, Hawthorn, VIC 3122, Australia}
\begin{document}

\date{draft \today}
%\pagerange{\pageref{firstpage}--\pageref{lastpage}}\pubyear{2012}
\maketitle

\begin{abstract}
While tens or hundreds of stellar-remnant black holes are expected to form in globular star clusters, it is still unclear how many of those will be retained upon formation, and how many will be ejected through subsequent dynamical interactions. No such black holes have been found in any Milky Way globular cluster until the recent discovery of stellar-mass black holes in the globular cluster M$22$ (NGC $6656$) with now an estimated population of $5-100$ black holes. We present a direct $N$-body model of a star cluster of the same absolute and dynamical age as M$22$. Imposing an initial retention fraction of $\approx 10\%$ for black holes, $16$ stellar-remnant black holes are retained at a cluster age of $12$ Gyr, in agreement with the estimate for M$22$. Of those $16$ BHs, two are in a binary system with a main sequence star each while also one pure black hole binary is present. We argue that multiple black holes can be present in any Milky Way cluster with an extended core radius, such as M$22$ or the model presented here.
\end{abstract}


%-----------------------------------------------------------------------

\section{INTRODUCTION}\label{intro}
Stellar-mass black holes (BHs) are formed at the endpoints of stellar evolution of very massive stars. Depending on the chemical composition or metallicity of a star, a supernova resulting in either a neutron star or a black hole is the ultimate fate for most stars above $\approx 7-8\,M_{\odot}$ \citep{Pols1998}. 

Tens or hundreds of such stellar-remnant BHs are expected to form in globular clusters (GCs), however at the current stage it is still not entirely clear how many will receive a velocity kick at formation and get ejected immediately, and how many BHs might be removed during the subsequent dyamical evolution of the cluster. While several extragalactic GCs containing BHs are known at the current stage (e.g. \citealt{Maccarone2011}), none had been confirmed in a Milky Way GC until the recent discovery of two such stellar-mass BHs in M$22$ (NGC $6656$, \citealt{Strader2012}). Since only BHs currently undergoing observable accretion can be detected via radio or $X$-ray emission and \citet{Strader2012} estimate that $2-40\%$ of BHs are expected to become members of binary systems with observable accretion over $10$ Gyr, it is possible that a total population of $\approx 5-100$ BHs exists in M$22$.

Theoretical predictions in the past have lead to the assumption that well before a cluster age of $12$ Gyr all but up to four (possibly all) BHs should be ejected from the cluster core \citep{Sigurdsson1993}, or similarly that nearly all BHs should be ejected, with the possibility of remaining BHs capturing normal stars to form low-mass X-ray binaries in low-density environments \citep{Kulkarni1993}. \citet{Strader2012} claim their observations to be in contrast to such prior theoretical predictions. However it was not possible to test this in direct $N$-body models comparable to clusters like M$22$ until very recently as such models are computationally expensive. We note that multiple BHs have already been found to remain in $N$-body models with smaller numbers of stars (e.g. \citealt{Mackey2007, Mackey2008}) as well as in Monte Carlos simulations of globular cluster evolution (e.g. \citealt{Downing2010, Downing2012}), while BHs are ejected from the cluster in $N$-body models by \citet{Banerjee2010} where high numbers of BHs were added initially.

Based on the recent findings by \citet{Repetto2012} suggesting that BHs receive similar velocity kicks as neutron stars (NSs) upon formation, and the observations that not all NSs are expected to receive a high kick \citep{Pfahl2002}, we present a direct $N$-body model for an intermediate-mass globular cluster containing $262\,500$ stars in total and incorporating a retention fraction of $\approx 10\%$ for both BHs and NSs. The cluster is evolved using \texttt{NBODY6} \citep{Aarseth1999, Aarseth2003} in a Milky Way-like gravitational potential field. 

While the model presented in this paper is by no means an attempt at a direct model of M$22$, it is representative owing to comparable dynamical and absolute ages. We evolve the model up to an age of $20\,$Gyr, and focus our analysis at the age of $12\,$Gyr: the estimated age for many globular clusters in the Milky Way including M$22$ \citep{Salaris2002}. At this stage, $\approx 180\,000$ stars are still retained within the cluster (including $16$ BHs), implying that this is the largest direct $N$-body model of a globular cluster currently available (cf. \citealt{Baumgardt2003, Hurley2012}).  

%
%-----------------------------------------------
%

\section{SIMULATION METHOD \& CHOICE OF PARAMETERS}\label{method}
We use the direct $N$-body code \texttt{NBODY6} \citep{Aarseth1999, Aarseth2003, Nitadori2012} to evolve our model and chose a set-up similar to the clusters presented in \citet{Sippel2012}. 
We evolve a star cluster with $N_i=250\,000$ stellar systems and an initial binary fraction of $b_f=0.05$, implying that the initial numbers of stars is in fact $N=262\,500$ including $25\,000$ stars in $12\,500$ binary systems. The stars are drawn from the initial mass function presented by \citet{Kroupa1993} within the limits $0.1 \leq m \leq 50\,M_{\odot}$ and distributed according to a Plummer density profile \citep{Plummer1911, Aarseth1974}. The initial cluster mass adds up to $M_i=1.6 \times 10^5\,M_{\odot}$. The conversion from $N$-body units to physical scale sizes leaves the scale radius as a free parameter which is set to give an initial half-mass radius of $r_{50\%}=6.2\,$pc.
The cluster is orbiting a Milky Way like galactic potential consisting of three components: a point-mass bulge, an extended smooth disc \citep{Miyamoto1975} and a logarithmic dark matter halo \citep{Aarseth2003}. We chose a galactocentric distance of $d_{gc}=8.5\,$kpc. 
From King model fits \citep{King1966} we find an initial tidal radius of $r_t\approx 50\,$pc for the $N$-body model, while M$22$ has a smaller tidal radius of $r_t=27\,$pc \citep{Mackey2005} owing to the smaller galactocentric distance of $d_{gc}=4.9\,$pc. 

Within the Milky Way, globular clusters across the whole range of metallicities exist around the galactocentric distance used here and we chose a low to intermediate metallicity Z$=0.001$ (corresponding to [Fe/H]$\approx -1.3$). This metallicity is close to that of M$22$ [Fe/H]$=-1.7$ and stars within this metallicity range evolve in an almost identical fashion (as illustrated in \citealt{SSE, Hurley2004, Sippel2012}). Stars are evolved according to stellar and binary evolution algorithms \citep{SSE, BSE} implying that stellar collisions may occur and stellar remnants such as white dwarfs (WDs), NSs and BHs form. BHs form from progenitors with masses greater than $20\,M_{\odot}$, where the resulting remnants are within the mass range of $4-30\,M_{\odot}$ \citep{SSE, Belczynski2010}. These massive stars evolve rapidly with most BHs forming during the first $\approx 10\,$Myr of cluster evolution.

\subsection{Velocity kicks}
It is well established that some NSs will receive velocity kicks upon formation resulting from asymmetries in the star’s surface before collapse \citep{Pfahl2002}. To mimic a neutron star retention fraction similar to indications from observations of $\approx 10 \%$ (c.f. \citealt{Pfahl2002, Pfahl2003}) we adopt a random flat kick distribution between $0-100\,$km/s. Recent indications suggest that BHs should receive similar kicks to NSs \citep{Repetto2012}, as such we apply an identical kick distribution to BHs. For our model, the escape velocity is $v_{\rm{esc}}=\sqrt{2GM/r_{\rm{50\%}}}=14\,$km/s at the very beginning of cluster evolution ($M$ is the cluster mass, $r_{\rm{50\%}}$ the half-mass radius and $G$ the gravitational constant). At an age of $200\,$Myr, by which time all BHs and NSs will have safely formed, the escape velocity is $10.9\,$km/s owing to $r_{\rm{50\%}}=8.5\,$pc and $M=11.8 \times 10^4\,M_{\odot}$.

\begin{figure}
\centering
\includegraphics[width=0.47\textwidth]{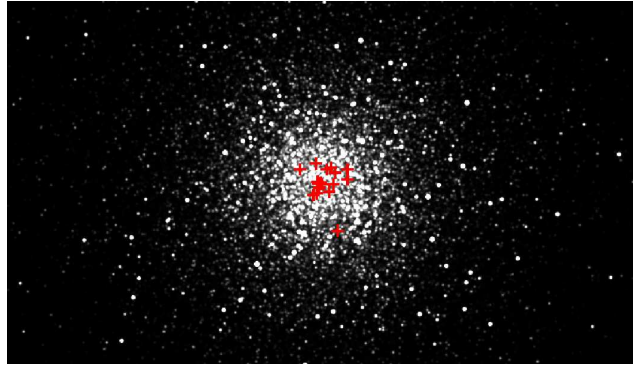}
\caption{Location of BHs (red crosses) at $12\,$Gyr of cluster age, projected in front of the cluster to increase visibility.}\label{fig:fig0}
\end{figure}

We note that in general, for a supernova occurring in a tight binary system, the kick velocity can be significantly above the cluster escape velocity and still result in one or both binary components being retained, with the binary loosened or broken up. Seven BHs form while part of a binary system, all of them being disrupted either immediately or within the next $\approx 10\,$Myr. 
Ultimately, this means that out of $\approx 450$ stellar remnant BHs formed through stellar evolution, $48$ are retained beyond the first $200\,$Myr, i.e. are not immediately ejected from the cluster via the velocity kick process. Within less than one Gyr, they have settled into the cluster core owing to mass-segregation and have a steadily decreasing half-mass radius of $\approx 1\,$pc. The spatial distribution of BHs within the model cluster at $12\,$Gyr is illustrated in Fig. \ref{fig:fig0} (see also Fig. \ref{fig:fig1}). 

\subsection{Size and time scales: measurements}
The sizes of $N$-body star cluster models are usually measured by either the half-mass radius or the core radius. Both quantities are measured in three dimensions, in contrast to observational size scales. The half-mass radius $r_{\rm{50\%}}$ is simply the $50\%$ Lagrangian radius, however the $N$-body core radius is a density weighted average distance from the cluster centre \citep{vonHoerner1960, vonHoerner1963, Casertano1985, Aarseth2003} and hence not comparable to the observational core radius resulting from e.g. King model fits to the surface brightness profile.
The procedure to calculate both observable half-light as well as core radii is similar to that described in \citet{Sippel2012}. In particular, we cross convolve the \texttt{NBODY6} output of mass, luminosity and radius for each single star with stellar atmosphere model calculations \citep{Kurucz1979} to obtain V-band magnitudes. Surface brightness profiles are produced by taking the average over projection along the $x-$, $y-$ and $z-$axes; and we use gridfit \citep{McLaughlin2008} to fit a King model to the cluster. 
From projection effects alone, the half-mass radius is $\approx 1.3$ times larger than the half-light radius.

%
%-----------------------------------------------------------------------
%

\section{EVOLUTION}\label{evolution}
\begin{figure}
\centering
\includegraphics[width=0.47\textwidth]{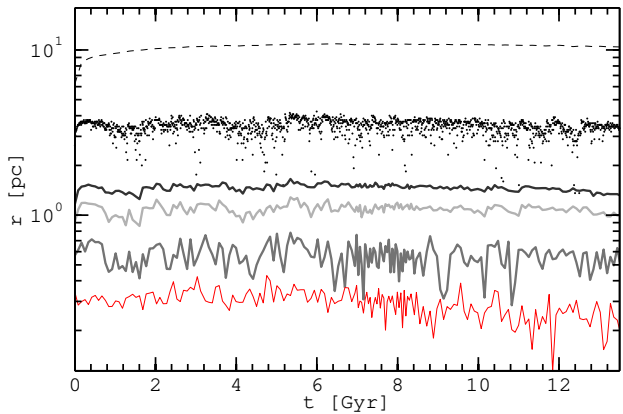}
\caption{Size evolution of the cluster over time, from top to bottom these are half-mass $r_{\rm{50\%}}$ (dashed), $N$-body core radius $r_{\rm{c}}$ (dotted), further Lagrangian radii $r_{\rm{1\%}}$, $r_{\rm{0.5\%}}$ and $r_{\rm{0.1\%}}$ as well as the half-mass radius for the BH population (red), always smaller than even the $0.1\%$ Lagrangian radius.}\label{fig:fig1}
\end{figure}

The first few Gyr of cluster evolution are dominated by heavy mass loss arising from stellar evolution as well as dynamical relaxation that results in an initial expansion of the cluster to a half-mass radius of up to $r_{\rm{50\%}} \approx 11\,$pc (Fig. \ref{fig:fig1}). The cluster then slowly starts to contract to $r_{\rm{50\%}}=10.6\,$pc and an $N$-body core radius $r_{tc}=3.1\,$pc, both three-dimensional quantities, at $12\,$Gyr. In terms of physical size, the model has a half light radius of $r_h=6.8\,$pc at $12\,$Gyr and we find an observational core radius of $r_{\rm{oc}}=3.6\,$pc. The half-mass relaxation time at $12\,$Gyr is $\approx 2.1\,$Gyr, where we utilize the half-\emph{light} radius to to calculate this number as suggested by \citet{Djorgovski1993} and \citet{Harris1996}. 

Owing to the influence of the external galactic tidal field as well as dynamical interactions, stars are continuously lost from the cluster, with $\approx 180\,000$ stars remaining at $12\,$Gyr. This corresponds to $\approx 68 \%$ with respect to the initial number of stars and a total mass of $M=6.7 \times 10^4 \,M_{\odot}$ ($\approx 42 \%$ of the initial mass).

Globular clusters are efficient in equalizing the distribution of energy, causing the most massive stars or remnants to sink to the centre, resulting in mass segregation. Even though our model is segregated after $12\,$Gyr, the average mass within the innermost pc is only $0.7\,M_{\odot}$ (compared to an overall average of $0.38\,M_{\odot}$). Simultaneously, heating of the core from either BHs or NSs can be expected to produce signatures such as an expanded core \citep{Mackey2007, Mackey2008}, where both effects can influence the surface brightness profile and are measurable by means of the core radius \citep{King1966}. M$22$ has a large core radius compared to other globular clusters in the Milky Way, indicating that a heating source may be present. While our model does not enter the phase of core collapse during it’s evolution, the subsystem of stars contained within the innermost $2\,$pc (dynamically dominated by BHs) is entering a phase of core-collapse at $\approx 1\,$Gyr of cluster age, indicated by the drop and fluctuation in the core radius (Fig. \ref{fig:fig1}). This results from the fact that the BH population has a much smaller half-mass relaxation time than the cluster as a whole, resulting the the BHs segregating to the cluster centre. The half-mass radius of the entire BH population is $\approx 0.2$ pc and smaller than the $0.1\%$ Lagrangian radius (see Fig. \ref{fig:fig1}). At $12\,$Gyr, the total mass contained in BHs is $\approx 190\,M_{\odot}$ corresponding to only $\approx 0.3\%$ of the overall cluster mass but $\approx 32 \%$ of the mass within the central parsec.

\begin{figure}
\centering
\includegraphics[width=0.47\textwidth]{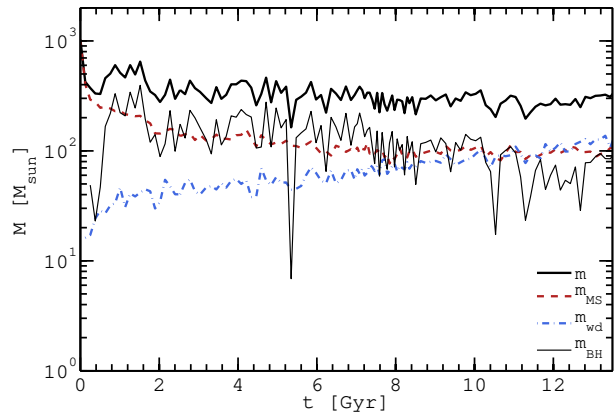}
\caption{Contribution of stars to the central parsec of the cluster, i.e. $d<1\,$pc, where the distance $d$ is again measured in three dimensions. It can be seen that at $12\,$Gyr, MS stars as well as WDs and BHs are contributing roughly $1/3$ of the dynamical mass each, while the contribution of other stars (giants or neutron stars) is minimal (i.e. $\leq 10\,M_{\odot}$ for both together). The ratio is different when looking at the number of stars in each category: at $12\,$Gyr, MS stars count for roughly $2/3$ of the stars within the innermost pc, while WDs make up $\approx 1/3$ of the stars. The dip in BH mass within the central region at $\approx 5.5\,$Gyr is caused by an energetic interaction within a close subsystem of three BHs, two of them previously in a BH-BH binary. One BH is ejected out of the cluster, while the BH-BH binary receives a recoil velocity, ejecting it momentarily beyond the central parsec. The loss of kinetic energy causes an expansion of the cluster core, however the BH-BH binary quickly sinks back to the centre.}\label{fig:fig2}
\end{figure}

This already indicates that BHs are an important component to the contribution of mass in the cluster centre. Indeed we find that at $12\,$Gyr, the innermost parsec (measured in three dimensions) is composed equally of main sequence stars, white dwarfs, and BHs (Fig. \ref{fig:fig2}). For the metallicity of both M$22$ and our model the main sequence turnoff mass is $0.83\,M_{\odot}$ at $12\,$Gyr, implying that most white dwarfs (with a peak in their mass distribution at $0.6\,M_{\odot}$) will be less massive than the main sequence turnoff, while all neutron stars and black holes will be more massive. In total, $245\,M_{\odot}$ is contained in the innermost pc in $352$ stars (compared to $2\,322$ stars or $1\,263\,M_{\odot}$ in projection), and this central density increases to $528$ stars or $423\,M_{\odot}$ at an age of $20\,$Gyr (or $2\,582$ stars and $1\,616\,M_{\odot}$ in projection).

\subsection{Binary systems with black holes}
Binary formation and disruption, as well as hardening of binary systems, are common processes in globular clusters. However it is easier to form new binary systems via exchange interactions than creating them in two-body encounters \citep{Heggie1975, HeggieHut}. A black hole can be part of a binary system where the other component can be at any stellar evolution stage. In our simulation, the first dynamical binary system including a BH forms at $\approx 380\,$Myr (a $25.3\,M_{\odot}$ BH and a $0.2\,M_{\odot}$ MS star). At $12\,$Gyr, the initial binary fraction of $b_f=0.05$ is slightly decreased to $b_f=0.045$ (with $7\,850$ binary systems present). In M$22$, one or potentially both observed BHs are in a binary with a low-mass main sequence star, or possibly a white dwarf \citep{Strader2012}. In our model, we find that at around $12\,$Gyr, two BHs are in a long-period binary system with a main sequence star each, where neither of those systems are undergoing mass transfer. Specifically, system a) consists of a BH with mass $6.2\,M_{\odot}$ and a main sequence star with mass $0.8\,M_{\odot}$ while system b) consists of a BH with mass $6.4\,M_{\odot}$ and a main sequence star of $0.4\,M_{\odot}$. Just before $12\,$Gyr a three-body encounter occurs involving system b) and a main sequence star with mass $0.7\,M_{\odot}$, replacing the lighter main sequence star and forming again a binary system consisting of a black hole and main sequence star. 

\begin{figure}
\centering
\includegraphics[width=0.47\textwidth]{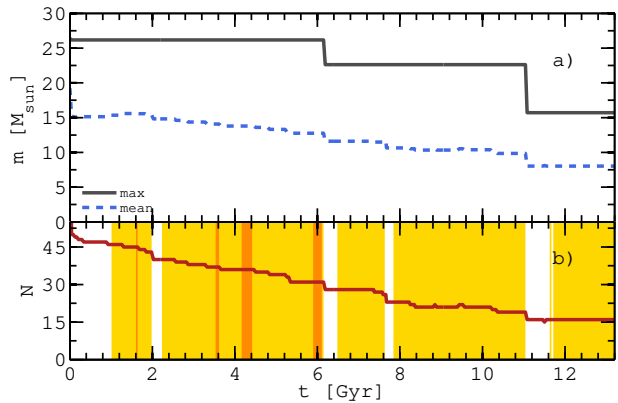}
\caption{Evolution of the mass of the entire black hole population a) and the number of BHs within the cluster b). In the upper panel, the solid black line denotes the maximum mass of the BHs still retained in the cluster, while the blue-dashed line denotes the mean mass of the BH population at a given time. The number of BHs is decreasing b) from $48$ at $200\,$Myr to $16$ BHs at $12\,$Gyr of cluster age. A yellow background denotes the presence of one BH-BH binary, while orange means two such systems are present, and white none. Note that the disruption of both BH-BH binary systems at $\approx 6\,$ and $\approx 11\,$Gyr is accompanied by the ejection of the most massive black hole contained in the cluster. The disruption of one BH-BH binary is also accompanied by the ejection of one BH at $\approx 2\,$ and $\approx 7.5\,$Gyr.}\label{fig:fig3}
\end{figure}

A BH-BH binary system is also present at around $12\,$Gyr, where the components of this binary have masses of $22\,M_{\odot}$ each. Just after $12\,$Gyr, during a three-body encounter with another BH with $13\,M_{\odot}$, this binary gets disrupted, whilst the individual components of the three-body system stay retained within the cluster. A BH-BH binary in a new combination forms quickly and the total number of $16$ BHs is stable around the $12\,$Gyr timeframe, as illustrated in  Fig. \ref{fig:fig3}. 
Dynamical interactions like these are what can ultimately cause the depletion of all but one BH or a tight BH-BH binary (which might ultimately merge, \citealt{Aarseth2012}), from the cluster. However it is not certain over which timescale the evaporation of black holes from globular clusters should take place. This depletion timescale is dependent on random encounters, necessary to eject BHs, while the frequency of such encounters depends on cluster parameters such as the core density \citep{Pooley2003}. We conclude that $12\,$Gyr is not enough time to deplete all or almost all BHs. In fact we have evolved the model up to $20\,$Gyr, and even at this late stage, $10$ BHs remain in the cluster. Five of those are in a binary system with either a main sequence star, white dwarf or another black hole. We also find a BH-NS system with a total mass of $\approx 18\,M_{\odot}$ at $\approx 18\,$Gyr in a stable configuration for $\approx 0.6\,$Gyr, at which stage the NS gets replaced by a BH with mass $7.1\,M_{\odot}$ in a three-body encounter. \citet{Clausen2012} have recently presented a study regarding BH-NS mergers in GCs and we conclude that dynamical interactions don't efficiently produce BH-NS binaries in GCs.

\subsection{Comparison}
While a direct model approach for M$22$ is still out of reach from a computational point of view, the $N$-body model presented in this work is comparable in many aspects. While our model has a half-mass relaxation time of $t_{\rm{rh}}=2.1\,$Gyr, the correspinding value is $t_{\rm{rh}}=1.7\,$Gyr for M$22$ \citep{Djorgovski1993, Harris1996}. M$22$ has a half-light $r_{\rm{h}}=3.1$ and core radius $r_{\rm{oc}}=1.2\,$pc \citep{Harris1996}, implying that the $N$-body model is less dense than M$22$. We argue that the higher central density of M$22$ does not severly change our predictions for stellar remnant BHs in globular clusters with large core radii since examples of denser model clusters with BHs exist in the literature. 

Most prominent is the $N$-body model by \citet{Hurley2012} consisting of $200\,000$ stars initially, evolved at a galactocentric distance of $3.9\,$kpc and with an initial half-mass radius of $4.7\,$pc. The strong gravitational field of the host galaxy at this distance has an impact on the cluster by stripping off stars in the outskirts, resulting in only $\approx 35\,000$ stars remaining at $12\,$Gyr. However, the half-mass relaxation time of the \citet{Hurley2012} model is much below that of M$22$ resulting in the model reaching the end of it’s initial phase of core-collapse at $11.5\,$Gyr, greatly exceeding the dynamical age of M$22$. At this stage, the core radius is $r_{\rm{oc}}=0.84\,$pc with four BHs still present (two of those in a BH-BH binary). This is the case even though the central density is significantly enhanced at this time (just over $3\,000$ stars in the innermost parsec, or $\approx 5\,600$ in projection). Even though the black holes ultimately get ejected during the evolution subsequent to core-collapse, we can safely conclude that BHs in the core of M$22$ can be expected, as the density of M$22$ lies in between the \citet{Hurley2012} model and the $N$-body model presented in this work -- however much closer to the new model here. 

Other examples of $N$-body cluster models with BHs currently exist in the literature (\citealt{Mackey2007, Mackey2008, Hurley2010} just to name a few), as well as Monte Carlo models of more massive globular clusters at various densities \citep{Downing2010, Downing2012, Morscher2012}. 
In addition we note that models involving mergers of stellar remnant black holes as well as well as intermediate mass BHs have been carried out in the past (e.g. \citealt{PortegiesZwart2002, Baumgardt2004, OLeary2006, Aarseth2012}).

\section{CONCLUSIONS}
We conclude that all three findings of \citet{Strader2012}: a) the observation of two stellar remnant black holes in M$22$ which are b) potentially in binary systems with main sequence stars and that c) $5-100$ stellar remnant black holes might be present in a cluster such as M$22$ -- are in excellent agreement with our direct $N$-body model. A key point in this analysis is that M$22$ has a large core radius compared to the average for the globular cluster population of the Milky Way, and is not \emph{dynamically} old. We conclude that multiple stellar remnant black holes in the core of such clusters can exist even beyond the current time. Furthermore, in the $N$-body model presented here we find that $1/3$ of the BHs retained in the cluster upon formation remain in the core beyond the Hubble time. 

With further advances in computing hardware we expect these results to be confirmed by even larger models with an expanded set of initial parameters in the near future.

\section{Acknowledgments}
We thank Juan Madrid, Marie Martig and Sverre Aarseth for comments and discussion. The $N$-body model was evolved for $\approx 250$ days on a NVIDIA Tesla C$1060$ graphics card at Swinburne University of Technology and the data analysis was carried out on the Green Machine at Swinburne. AS acknowledges a SUPRA scholarship from Swinburne University.
%---------------------------------------------------------------------
%---------------------------------------------------------------------
\bibliography{refs}
\end{document}